\documentclass[aps,prl,showpacs,twocolumn,amsmath,amssymb,superscriptaddress,letterpaper]{revtex4}
\usepackage{times}
\usepackage{amsfonts}
\usepackage{mathrsfs}
\usepackage{graphicx}
\usepackage{dcolumn}
\usepackage{bm}
\usepackage{color}

\usepackage[colorlinks,bookmarks=false,citecolor=blue,linkcolor=red,urlcolor=blue]{hyperref}
\def\be{\begin{equation}}       \def\ee{\end{equation}}
\def\bea{\begin{eqnarray}}      \def\eea{\end{eqnarray}}

\begin{document}
\title{Theoretical studies of superconductivity in doped  BaCoSO}

\author{Shengshan Qin}
\affiliation{Beijing National Laboratory for Condensed Matter Physics, and Institute of Physics, Chinese Academy of Sciences, Beijing 100190, China}

\author{Yinxiang Li}
\affiliation{Beijing National Laboratory for Condensed Matter Physics, and Institute of Physics, Chinese Academy of Sciences, Beijing 100190, China}

\author{Qiang Zhang}
\affiliation{Department of Physics and Astronomy, Purdue University, West Lafayette, Indiana 47907, USA}

\author{Congcong Le}
\affiliation{Beijing National Laboratory for Condensed Matter Physics, and Institute of Physics, Chinese Academy of Sciences, Beijing 100190, China}

\author{Jiangping Hu}\email{jphu@iphy.ac.cn }
 \affiliation{Beijing National Laboratory for Condensed Matter Physics, and Institute of Physics, Chinese Academy of Sciences, Beijing 100190, China}
\affiliation{Collaborative Innovation Center of Quantum Matter, Beijing, China}
\affiliation{Kavli Institute of Theoretical Sciences, University of Chinese Academy of Sciences,  Beijing 100049, China}

\date{\today}

\begin{abstract}
We investigate  superconductivity that may exist in the doped BaCoSO, a multi-orbital Mott insulator with a strong antiferromagnetic ground state.  The superconductivity is studied in both t-J type and Hubbard type multi-orbital models  by  mean field approach and random phase approximation (RPA) analysis. Even if there is no $C_4$ rotational symmetry, it is found that  the system still carries  a d-wave like pairing symmetry state with gapless nodes and  sign changed superconducting order parameters on Fermi surfaces.  The results are largely doping insensitive.   In this superconducting state, the three $t_{2g}$ orbitals have very different superconducting form factors in momentum space.  In particular,    the intra-orbital pairing of the $d_{x^2-y^2}$  orbital has a s-wave like pairing form factor.  The two methods also  predict very different pairing strength on different parts of Fermi surfaces.These results suggest that BaCoSO and related materials can be a new ground to test and establish fundamental principles for unconventional  high temperature superconductivity.

\end{abstract}


\maketitle

\section{Introduction}

Since the discovery of cuprates\cite{cuprates} and the iron based high T$_c$ superconductors\cite{iron_based} (SCs), intensive research efforts have been made to understand their superconducting pairing mechanism. In the past three decades, great progress has  been made both experimentally and theoretically. Various experimental techniques, such as angle-resolved photoemission spectroscopy \cite{ARPES1} (ARPES), inelastic neutron scattering\cite{INS2} (INS), scanning tunneling spectroscopy\cite{STS2} (STS), etc. and many different theories have been developed in the research of high T$_c$ SCs. However, due to the complexity of the problem, no consensus about the microscopic pairing mechanism has been reached.

Recently, by comparing cuprates and the iron based SCs,  we have pointed out that  those $d$-orbitals that are responsible for the superexchange antiferromagnetic (AFM) interactions mediated through anions  are isolated near Fermi energy to generate superconductivity in both families of high T$_c$ SCs\cite{high_Tc_gene}.   In this scenario,  the pairing symmetry can be simply determined through an emergent empirical principle,  the  Hu-Ding principle\cite{hu_ding}.    More interestingly, this  electronic feature is largely absent in other correlated electron systems. Thus,  we have suggested that  this property can be the gene of unconventional high T$_c$ SCs and   materials satisfying the condition can be  promising high T$_c$ candidates.  Based on such an understanding, two families of materials\cite{hu1,hu2} have been proposed to be promising high T$_c$ SCs. However, the proposals have  not been tested until now because of the difficulty in synthesizing the proposed materials.

However,  recently we have observed that an already-synthesized material BaCoSO\cite{BaCoSO_AFM1,BaCoSO_AFM2} may give us a chance to test the theory. The lattice structure of BaCoSO is similar to the case in ref.\cite{hu2}, but the tetrahedron environment around the Co atoms is broken because of the anion mixture of the O and S atoms. Though it is not the ideal structure to maintain the required electronic condition,  we have expected that the theory\cite{high_Tc_gene} is  suitable to BaCoSO  and superconductivity may arise in the doped BaCoSO \cite{BaCoSO_hu} if the structure distortion and disorder induced by doping can be minimized.   Motivated by this, we carry out the theoretical investigation of the superconducting state in this type of electronic structures.

The paper is organized as follows. In the first part, the electronic structure and magnetic property of BaCoSO are reviewed. In the second part, the superconducting pairing in doped BaCoSO is analyzed by the mean field theory based on a t-J type multi-orbital model.  In the third part, we study the superconducting state in a Hubbard type of model under the RPA approximation. Then, we come to our conclusion in the last part.

\section{electronic structure and magnetic property}
BaCoSO has an orthorhombic layered lattice structure in which each CoSO layer is constructed by vertex sharing mixed-anion tetrahedron complexes CoS$_2$O$_2$. In this layer structure, the Co chains along $y$-direction are connected through S atoms and the staggered Co chains along $x$-direction are linked through O atoms. Compared with a perfect tetrahedron environment in which  the crystal field splits the five $d$-orbitals into two groups, $t_{2g}$ and $e_{g}$,  the  crystal field  here breaks  the degeneracy of the three $t_{2g}$ orbitals as well.  However, we have shown that this breaking is relatively small and the three $t_{2g}$ orbitals still control the major electronic physics.  Without doping, BaCoSO has been confirmed to be an antiferromagnetic(AFM)  Mott insulator both theoretically\cite{BaCoSO_hu} and experimentally\cite{BaCoSO_AFM1,BaCoSO_AFM2}.   The  AFM  order is G-type, similar to those of cuprates.

\begin{figure}
\centerline{\includegraphics[width=0.45\textwidth]{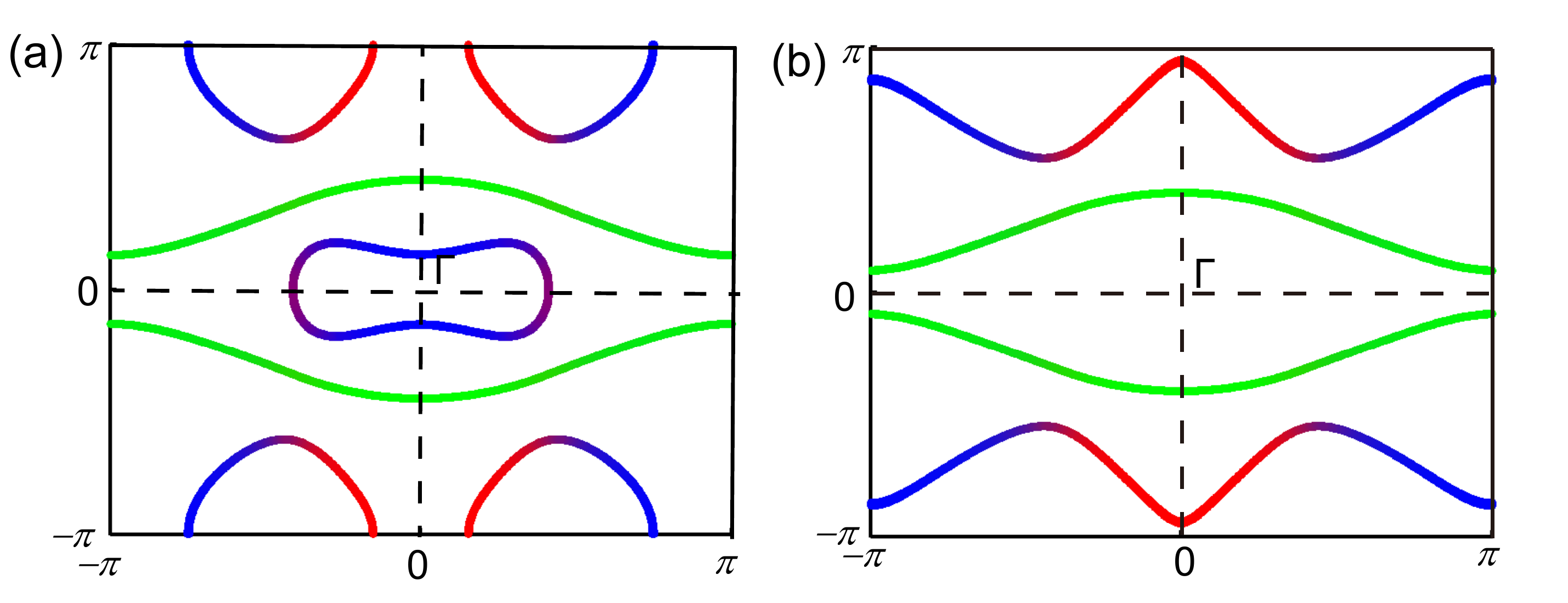}}
\caption{(color online) The FSs in the unfolded BZ based on the three band model are shown in (a) when the electron doping is $0.6$ per site and (b) when the electron doping is near $1.0$ per site. The orbital contributions of the different FS sheets are shown color coded: $d_{xz}$ (red), $d_{yz}$ (green) and $d_{x^2-y^2}$ (blue).
\label{FS}}
\end{figure}

With doping, the electronic structure in the normal state has been calculated in ref.\cite{BaCoSO_hu}.  It has been shown that  the three $t_{2g}$ orbitals dominate near the Fermi level. In this paper, we investigate the possible superconducting state of this system and  perform calculations based on the three band model in the unfolded Brillouin zone (BZ) constructed by the three $t_{2g}$ orbitals ($d_{xz}$,$d_{yz}$,$d_{x^2-y^2}$) in ref.\cite{BaCoSO_hu}

\begin{eqnarray}
\label{Hamiltonian_three_band}
H_0^{11}&=& \epsilon_{1}+2t^{11}_{xx}cos(2k_x) +2t^{11}_{y}cos(k_y)+4t^{11}_{yy}cos(2k_y) \nonumber\\
&&-2t^{11}_{x}cos(k_x)-4t^{11}_{xy}cos(k_x)cos(k_y), \nonumber\\
H_0^{12}&=&4t^{12}_{xy}sin(k_x)sin(k_y), \nonumber\\
H_0^{13}&=& 2it^{13}_{xx}sin(2k_x)+2it^{13}_{x}sin(k_x)+4it^{13}_{xy}sin(k_x)cos(k_y), \nonumber\\
H_0^{22}&=& \epsilon_{2}+2t^{22}_{xx}cos(2k_x)+2t^{22}_{y}cos(k_y)+2t^{22}_{yy}cos(2k_y)\nonumber\\
&&-2t^{22}_{x}cos(k_x)-4t^{22}_{xy}cos(k_x)cos(k_y), \nonumber\\
H_0^{23}&=& 2it^{23}_{y}sin(k_y)+2it^{23}_{yy}sin(2k_y)+4it^{23}_{xy}cos(k_x)sin(k_y), \nonumber\\
H_0^{33}&=& \epsilon_{3}+2t^{33}_{xx}cos(2k_x)+2t^{33}_{y}cos(k_y)+2t^{33}_{yy}cos(2k_y)\nonumber\\
&&+2t^{33}_{x}cos(k_x)+4t^{33}_{xy}cos(k_x)cos(k_y),
\end{eqnarray}

the corresponding tight binding parameters are shown in Table.\ref{3_orbital}.
The three bands model captures  the major electronic structure around the Fermi energy.  In Fig.\ref{FS}, we plot Fermi surfaces and their corresponding orbital characters at two different electron doping levels. For the case of 0.6 electron doping away from the half filling,  the Fermi surfaces (FSs) as shown in Fig.\ref{FS}(a) are composed of three pockets. A small hole pocket at the BZ center $\Gamma$ point  is attributed to the $d_{x^2-y^2}$ and $d_{xz}$ orbitals, so do the two electron pockets around  the BZ  boundary Y point.  A large open hole pocket around the BZ center is attributed to the $d_{yz}$ orbital.  The hybridization of  the $d_{x^2-y^2}$ and $d_{xz}$ orbitals stems from the the Zig-Zag Co chain structure along the x direction.  With heavy electron doping, the small hole pocket at  the BZ center can sink below the Fermi energy as shown in Fig.\ref{FS}(b).

\begin{table}[bt]
\caption{\label{3_orbital}%
The hopping parameters $t^{mn}_i$ between different neighbours for the three orbitals tight binding model\cite{BaCoSO_hu} for monolayer BaCoSO. The on-site energies of the three $t_{2g}$ orbitals are (all in eV): $\epsilon_{1}$ = -0.405, $\epsilon_{2}$ = -0.507, $\epsilon_{3}$ =-0.178.}
\begin{ruledtabular}
\begin{tabular}{cccccc}

  $t^{mn}_i$ & i=x &i=xx & i=y & i=yy & i=xy \\
 \colrule
mn=11 & -0.323 & 0.051 & 0.207  & -0.012 & -0.014 \\
mn=12 &  &  &  &  & 0.025 \\
mn=13 & 0.137 & -0.002 &  &  & 0.033 \\
mn=22 & -0.204 & -0.014 & 0.412 & 0.077 & -0.003 \\
mn=23 &  &  & 0.093 & 0.012 & -0.051 \\
mn=33 & -0.225 & -0.028 & 0.22 & 0.033 & 0.026 \\
\end{tabular}
\end{ruledtabular}
\end{table}

The above electronic structure resembles those of iron-based SCs. We can make a good comparison between them. In  iron-based SCs, typically there are  also three types of pockets,   two hole pockets from $d_{xz,yz}$  and  one hole pocket from $d_{xy}$ at the BZ center $\Gamma$ point, and two electron pockets around the BZ corner at $M$ point that are mixed with $d_{xy}$ and $d_{xz,yz}$\cite{Kuroki,Johnston}.  The hole pockets at $\Gamma$ point in iron-based SCs can also sink below FSs by electron doping\cite{Kuroki}. The role of different pockets and the interactions among them have been the central debate in iron-based SCs\cite{hu_ding,Mazin,Hirschfeld}. Therefore, the study of this new material can provide much deep understanding on these issues.

\section{mean field analysis}

We start with an effective t-J type Hamiltonian for BaCoSO, which is generally written as
\begin{eqnarray}
H= \tilde{H}_0+\sum_{<ij>,\alpha,a,b} (J^\alpha_{ab}  \vec S_{ia}\cdot\vec S_{jb}-\frac{1}{4}n_{ia}n_{jb}),
\label{Htj}
\end{eqnarray}
where $\tilde{H}_0$ is  the three bands Hamiltonian in Eq.\ref{Hamiltonian_three_band}\cite{BaCoSO_hu} subject to a projection to non-double occupant orbital state due to the onsite Hubbard interaction, $<ij>$ labels the two nearest neighbour (NN) sites, $\alpha=x,y$ labels direction and $a, b$ are the orbital indexes.  In local atomic orbital approximation, the strength of the AFM interaction can be roughly estimated from the superexchange process.  $J^\alpha_{ab}$ takes the form
\begin{eqnarray}
J^\alpha_{ab}=(t^\alpha_{ab})^2(\frac{1}{U_d}+\frac{1}{U_d+\Delta_{pd}}),
\end{eqnarray}
where t$^\alpha_{ab}$ is the effective hoping parameter between the NN $d$-orbitals at Co sites, $U_d$ is the Coulomb interaction for the $d$-orbitals and $\Delta_{pd}$ is defined as the energy difference between the $d$-orbital and $p$-orbital at O and S atoms. With the parameters in ref.\cite{BaCoSO_hu}, we can get the AFM interaction strength in BaCoSO as follows: $J^x_{xz}=0.20$eV, $J^x_{yz}=0.09$eV, $J^x_{x^2-y^2}=0.15$eV and $J^y_{xz}=0.09$eV, $J^y_{yz}=0.36 $eV, $J^y_{x^2-y^2}=0.09$eV.  It is important to note that this estimation is entirely based on atomic orbitals. As the $p$-orbitals of O and S atoms are very different  and the latter are more extended than the former,   the effective AFM coupling through S atoms is expected to be smaller than the estimated values. For this reason, we set a variable $\beta$ on the value of $\tilde{J^y}=\beta J^y$.

\begin{figure}
\centerline{\includegraphics[width=0.45\textwidth]{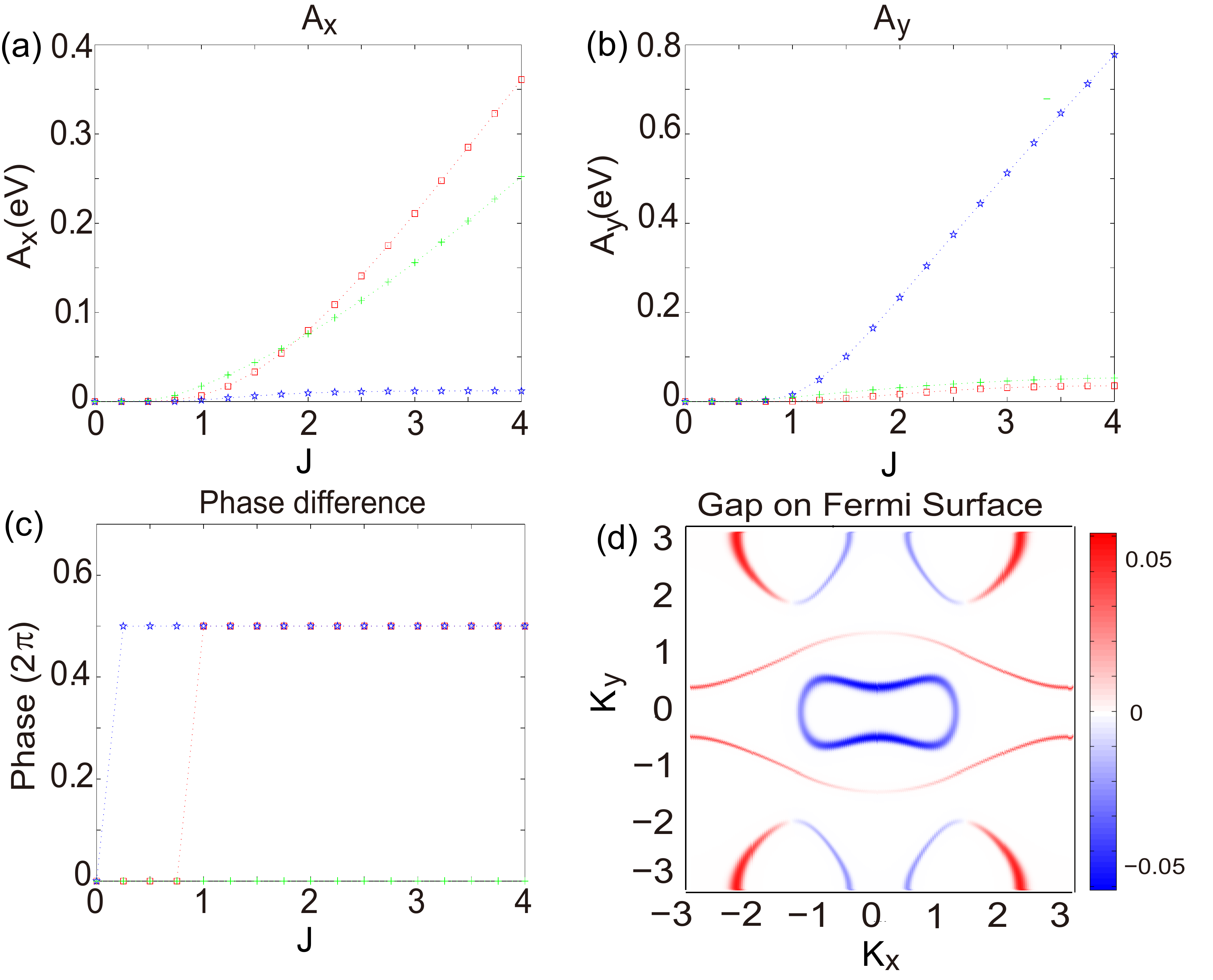}}
\caption{(color online) Pairing strength and superconducting gap on the FSs when the electron doping is $0.6$ per site. (a) and (b) show the pairing amplitude for different orbitals in the $x$-direction and the $y$-direction, respectively. The pairing phase difference in two directions is shown in (c). (d) shows the superconducting gap on the FSs when $J=\beta=1$, and in this case the order parameters are $\Delta^x_{xz}=-6.6$meV, $\Delta^x_{yz}=-1.6$meV, $\Delta^x_{x^2-y^2}=-17.2$meV and $\Delta^y_{xz}=0.8$meV, $\Delta^y_{yz}=14.8$meV, $\Delta^y_{x^2-y^2}=-9.7$meV, correspondingly. The amplitude and phase difference of the superconducting order parameters for different orbitals are shown as: $d_{xz}$ (red square), $d_{yz}$ (blue star) and $d_{x^2-y^2}$ (green plus).
\label{meanfield_low}}
\end{figure}

Because of the space anisotropy of the two $d$-orbitals, the AFM interactions  for the  $d_{xz}$ and $d_{yz}$ orbitals  and those for the $d_{x^2-y^2}$ orbital along the two different directions have significant anisotropy. Such a large anisotropy suggests that the electronic physics here is rather nematic.

 In the mean field calculation, we  approximate   the projection as an overall renormalization factor of the bare Hamiltonian, namely $\tilde{H}_0=\gamma H_0$, where $\gamma$ is the renormalization factor\cite{mean_feild1,mean_feild2}.  $\gamma$ generally is doping dependent and can be measured experimentally.  By rescaling the energy,  it is also equivalent to absorb the renormalization factor into the interaction parameters so that we can simply treat  $\tilde{H}_0=  H_0$ in the mean field calculation.  Combining with the estimated bare AFM interaction parameters, the mean field calculation is performed by setting AFM interaction in the unit of eV for the corresponding orbitals to be $[0.20, 0.09, 0.15]\times J$  along the $x$-direction and $[0.09, 0.36, 0.09]\times \beta J$ along the $y$-direction.  We report the phase diagram of the superconducting  state with respect to  $\beta$ and $J$.

\begin{figure}
\centerline{\includegraphics[width=0.45\textwidth]{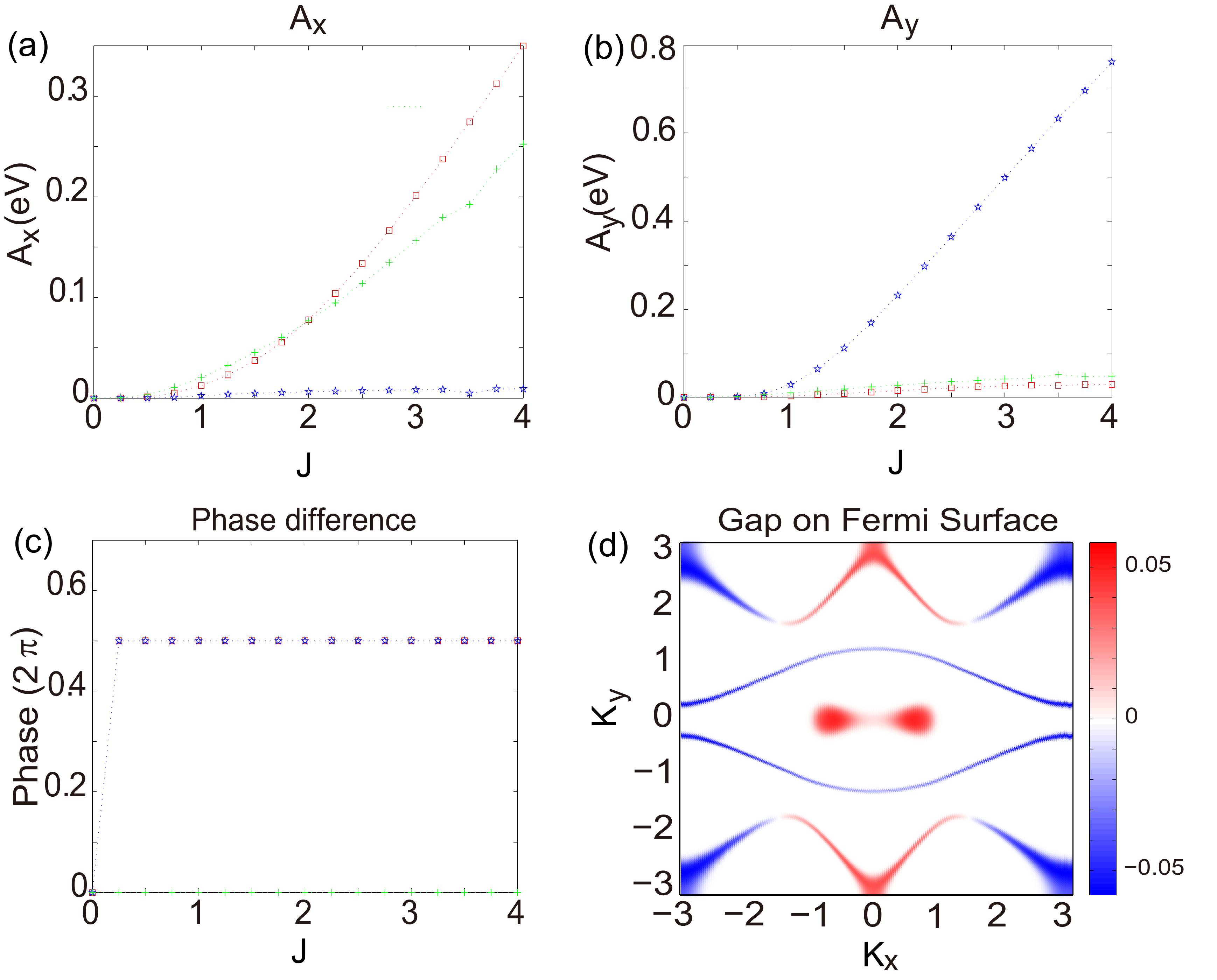}}
\caption{(color online) Pairing strength and superconducting gap on the FSs when the electron doping is about $1.0$ per site. (a) and (b) show the pairing amplitude for different orbitals in the x-direction and the y-direction, respectively. The pairing phase difference in two directions is shown  in (c). (d) shows the superconducting gap on the FSs when $J=\beta=1$, and in this case the order parameters are $\Delta^x_{xz}=12$meV, $\Delta^x_{yz}=2$meV, $\Delta^x_{x^2-y^2}=21$meV and $\Delta^y_{xz}=-3$meV, $\Delta^y_{yz}=-28$meV, $\Delta^y_{x^2-y^2}=10$meV, correspondingly. The amplitude and phase difference of  the superconducting order parameters for different orbitals are shown as: $d_{xz}$ (red square), $d_{yz}$ (blue star) and $d_{x^2-y^2}$ (green plus).
\label{meanfield_heavy}}
\end{figure}

First, we set $\beta=1$ and report results as a function of $J$  for two different doping levels. The FSs is shown in Fig.\ref{FS}(a) when the doping level is $0.6$ electron per site, and the corresponding mean field results are shown in Fig.\ref{meanfield_low}. The significant anisotropy of the AFM interaction for different orbitals leads to that, the pairing in the $y$-direction is dominated by $d_{yz}$ orbital while in the $x$-direction $d_{xz}$ and $d_{x^2-y^2}$ orbitals are dominant. As is shown in Fig.\ref{meanfield_low}(c), the superconducting order parameter for $d_{x^2-y^2}$ is s-wave like, while it is d-wave like for both $d_{xz}$ and $d_{yz}$ orbitals. Furthermore, the following relationship is satisfied in the main area of the  phase diagram

\begin{eqnarray}\label{relative phase}
sign(\Delta^x_{xz})=-sign(\Delta^y_{yz}),\nonumber\\
sign(\Delta^x_{xz})=sign(\Delta^x_{x^2-y^2}),
\end{eqnarray}

The results of the mean field approach can be well understood within the Hu-Ding principle\cite{hu_ding}, the superconducting ground state always tends to open the largest superconducting gap on the FSs. For the sake that only the NN AFM interaction is considered, the superconducting order parameter takes a form factor in momentum space as $\Delta^xcos k_x+\Delta^ycos k_y$ and $\Delta^\alpha$ is proportional to $J^\alpha$. Meanwhile, the three $t_{2g}$ orbitals hybridize only at several small area on the FSs, as is shown in Fig.\ref{FS}(a). Therefore, it is easy to see that   a s-wave like form factor $cos k_x+cos k_y$  for the $d_{x^2-y^2}$ orbital and a  d-wave like pairing on $d_{xz}$ and $d_{yz}$ orbitals can open the largest gaps on the FSs. The phase relationship of the order parameters between different orbitals can be also determined at the area where different orbitals hybridize on the FSs. To achieve  larger superconducting gaps, the order parameters of these orbitals tend to have the same phase. Specifically, for $0.6$ electron doped BaCoSO, because the $d_{yz}$ orbital and $d_{x^2-y^2}$ orbital hybridize strongly at the smaller FS near the $\Gamma$ point, $\Delta^x_{xz}$ tends to have the same phase with $\Delta^x_{x^2-y^2}$.

\begin{figure}
\centerline{\includegraphics[width=0.45\textwidth]{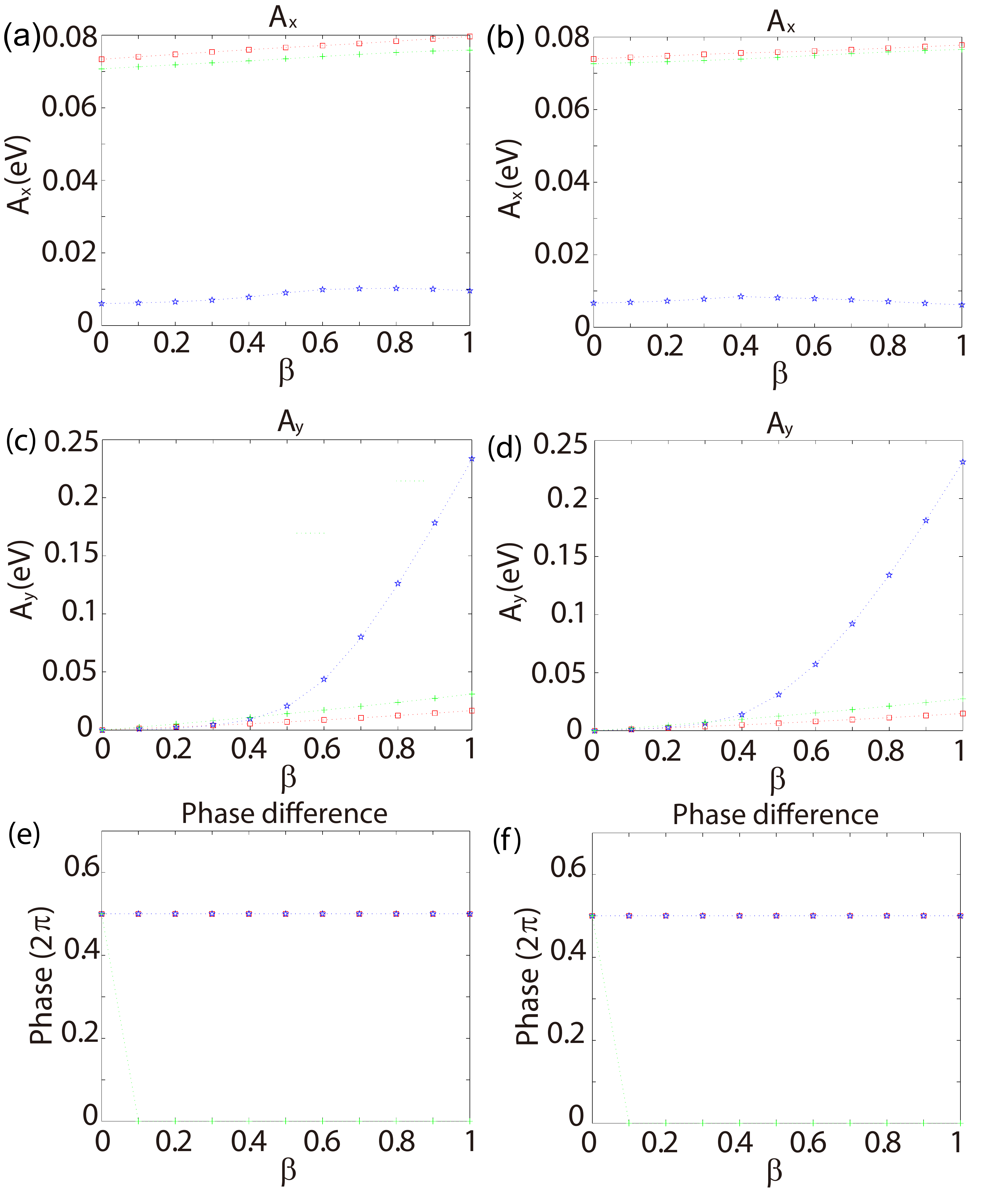}}
\caption{(color online) Pairing strength versus $\beta$ for different doping levels. (a) (c) show the pairing amplitude for different orbitals in the x-direction and the y-direction and (e) shows the pairing phase difference in the two directions, respectively, for the doping level which is about $0.6$ electron per site. (b) (d) and (f) show that for the doping level which is about $1.0$ electron per site, correspondingly. The amplitude and phase difference of the superconducting order parameters for different orbitals are shown as: $d_{xz}$ (red square), $d_{yz}$ (blue star) and $d_{x^2-y^2}$ (green plus).
\label{meanfield_beta}}
\end{figure}

A similar mean field analysis is also done when the electron doping level is about $1.0$ per site. The corresponding FSs and mean field results are shown in Fig.\ref{FS}(b) and Fig.\ref{meanfield_heavy}, respectively. The mean field results here are  similar to those of  the $0.6$ electron doped case. Furthermore,  we set $J=2.0$ and report results as a function of $\beta$. As shown in Fig. \ref{meanfield_beta}, the results are  also similar.   The qualitative results on  the superconducting order parameters  are very robust against $\beta$.

Overall, the mean field theory gives a rather robust superconducting state: a s-wave like order parameter for $d_{x^2-y^2}$ orbital, a d-wave like order parameter for both $d_{xz,yz}$   orbitals, and totally a d-wave like pairing symmetry on the FSs with nodes near $(\frac{\pi}{2},\frac{\pi}{2})$.

\section{random phase approximation analysis}

Based on the three bands model above, the RPA analysis is carried out for BaCoSO in this section with onsite repulsive interactions. The total Hamiltonian   is given by
\begin{equation}
\begin{aligned}
\emph{H}=&\emph{H}_{0}+U\sum_{i,\alpha}n_{i\alpha\uparrow}n_{i\alpha\downarrow}+U^{'}\sum_{i,\alpha<\beta}n_{i\alpha}n_{i\beta}\\
         &+J\sum_{i,\alpha<\beta,\sigma\sigma^{'}}c^{\dagger}_{i\alpha\sigma}c^{\dagger}_{i\beta\sigma^{,}}c_{i\alpha\sigma^{'}}c_{i\beta\sigma}\\
         &+J^{'}\sum_{i,\alpha\neq\beta}c^{\dagger}_{i\alpha\uparrow}c^{\dagger}_{i\alpha\downarrow}c_{i\beta\downarrow}c_{i\beta\uparrow},
\end{aligned}
\end{equation}
where  $n_{i,\alpha}=n_{i,\alpha,\uparrow}+n_{i,\alpha,\downarrow}$. For other indexes, we adopt the parameter notations given in ref.\cite{Kemper}. In the RPA approximation, the pairing vertex is

\begin{figure}
\centerline{\includegraphics[width=0.45\textwidth]{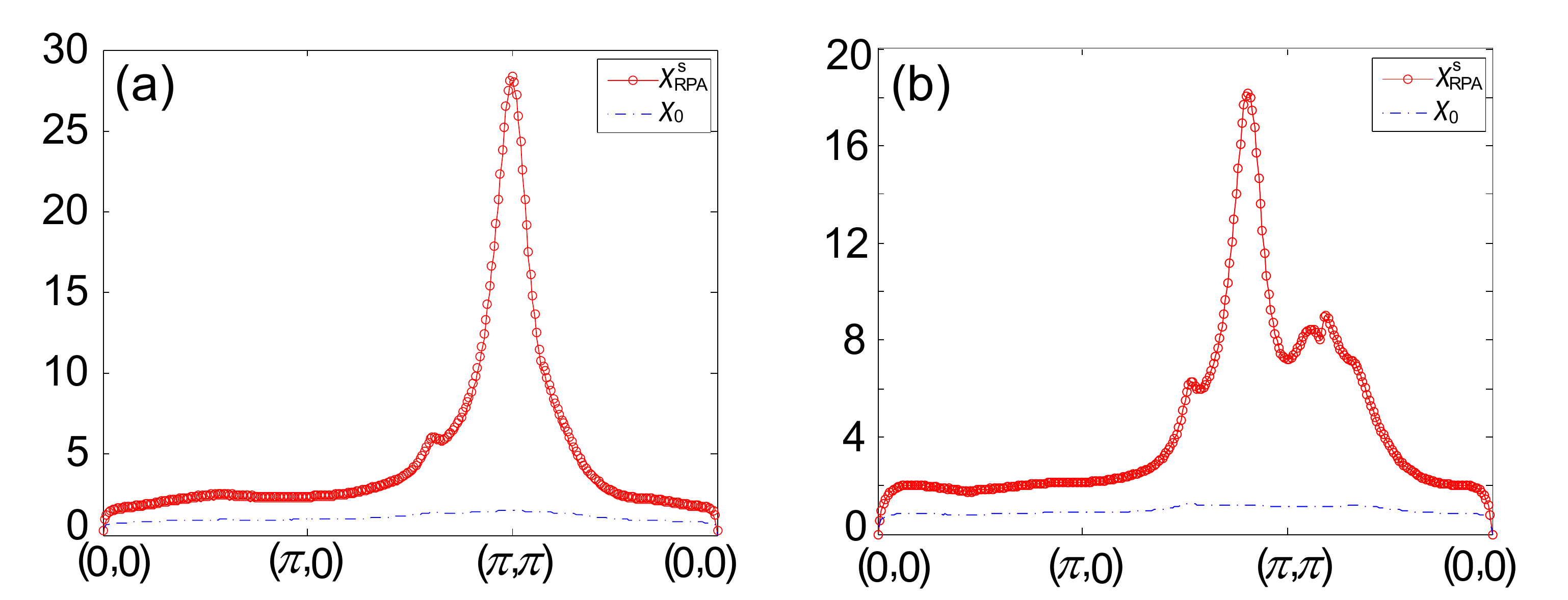}}
\caption{(color online) The real part of the bare (blue dot dash line) and RPA (red dot line) spin susceptibilities for BaCoSO for $0.6$ electron doping per site in (a) and $1.0$ electron doping per site in (b) along the high symmetry line.
\label{susceptibility}}
\end{figure}

\begin{equation}
\begin{aligned}
\Gamma_{ij}(k,k^{'})=&\emph{Re}\bigg[\sum_{l_{1}l_{2}l_{3}l_{4}}a^{l_{2},\ast}_{\emph{v}_{i}}(k)a^{l_{3},\ast}_{\emph{v}_{i}}(-k)\\
&\times\Gamma_{l_{1}l_{2}l_{3}l_{4}}(k,k^{'},\omega=0)a^{l_{1}}_{\emph{v}_{j}}(k^{'})a^{l_{4}}_{\emph{v}_{j}}(-k^{'})\bigg],
\end{aligned}
\end{equation}
where the momenta $k$ and $k^{'}$ is restricted to different FSs within an energy cutoff $\Lambda$,  with $k\in C_{i}$ and $k^{'}\in C_{j}$. $a^{l}_{v}$(orbital index $l$ and band index $v$) is the component of the eigenvectors of the three-orbitals tight binding Hamiltonian.
The singlet channel of orbital vertex function $\Gamma_{l_{1}l_{2}l_{3}l_{4}}$ in RPA is given by

\begin{equation}
\begin{aligned}
\Gamma_{l_{1}l_{2}l_{3}l_{4}}(k,k^{'},\omega)=&\bigg[\frac{3}{2}\bar{U}^{s}\chi^{RPA}_{1}(k-k^{'},\omega)\bar{U}^{s}+\frac{1}{2}\bar{U}^{s}\\
&-\frac{1}{2}\bar{U}^{c}\chi^{RPA}_{0}(k-k^{'},\omega)\bar{U}^{c}+\frac{1}{2}\bar{U}^{c}\bigg]_{l_{3}l_{4}l_{1}l_{2}},
\end{aligned}
\end{equation}
where $\chi^{RPA}_{1}$ and $\chi^{RPA}_{0}$ are the spin and charge fluctuation RPA susceptibility, respectively. The spin and charge interaction matrix($\bar{U}^{s}$, $\bar{U}^{c}$) are the same as in ref.\cite{Kemper}.
The pairing strength function is

\begin{eqnarray}\label{RPA_pair}
\lambda\big[\emph{g}(k)\big]=-\frac{\sum_{ij}\oint_{C_{i}}\frac{d\emph{k}_{\|}}{\emph{v}_{\emph{F}}(k)}\oint_{C_{j}}\frac{d\emph{k}^{'}_{\|}}{\emph{v}_{\emph{F}}(k^{'})}\emph{g}(k)\Gamma_{ij}(k,k^{'})\emph{g}(k^{'})}{(2\pi)^{2}\sum_{i}\oint_{C_{i}}\frac{d\emph{k}_{\|}}{\emph{v}_{\emph{F}}(k)}\big[\emph{g}(k)\big]^{2}},
\end{eqnarray}
where $v_{F}(k)=|\nabla_{k}E_{i}(k)|$ is the Fermi velocity on a given Fermi surface sheet $C_{i}$.
The calculation is carried out in the spin-rotational invariance case meaning $\bar{U^{'}}=\bar{U}-2J$ and $J=J^{'}$.

\begin{figure}
\centerline{\includegraphics[width=0.45\textwidth]{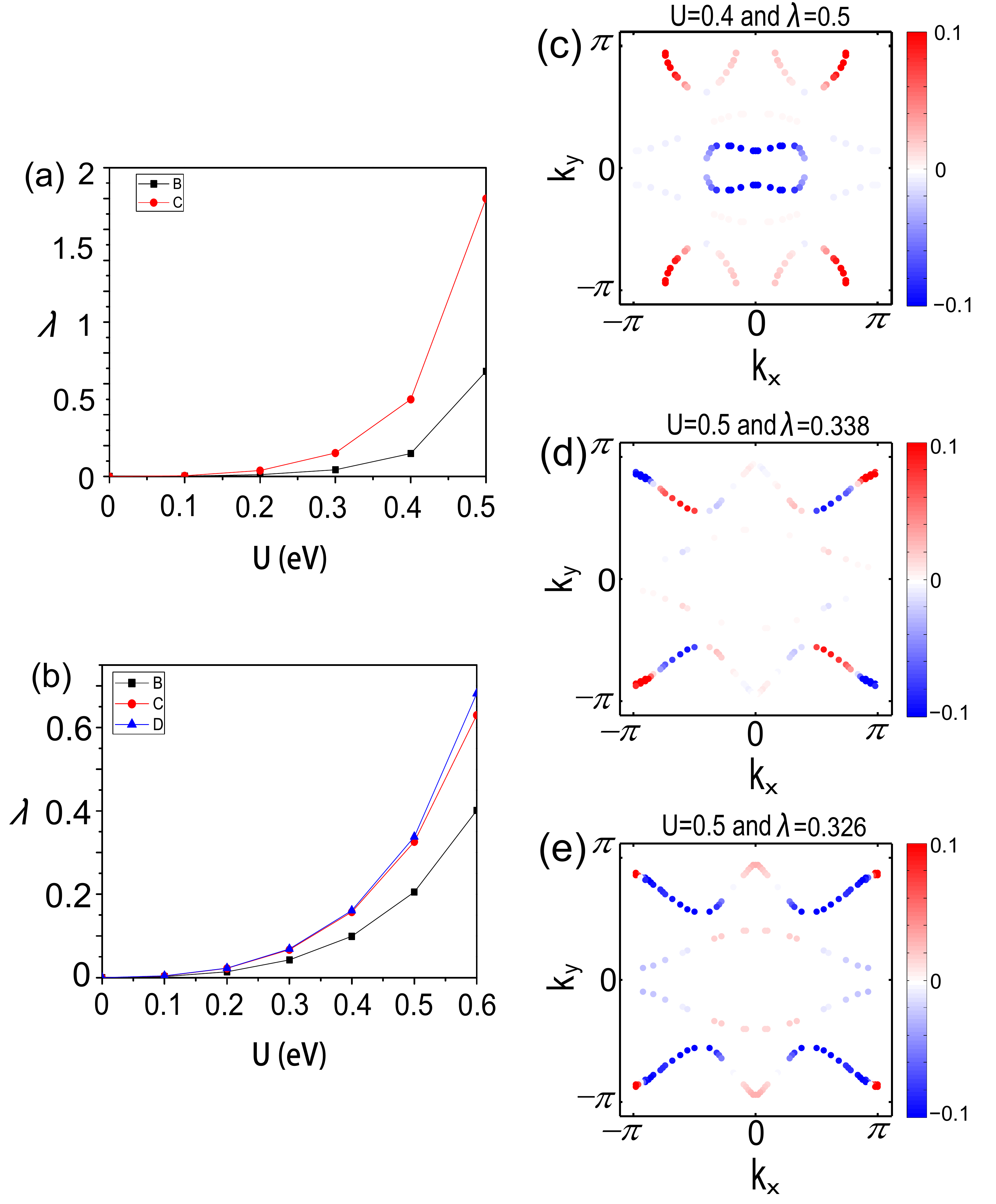}}
\caption{(color online) Pairing strength $\lambda$ and superconducting gap on the FSs of BaCoSO for $J/U=0.2$. (a) and (c) show the leading pairing instability and superconducting gap on the FSs when it is $0.6$ electron doped per site, respectively. (b) (d) and (e) show that for the $1.0$ electron doping per site case, correspondingly. For the heavy electron doping case, the pairing instabilities in (d) and (e) are nearly degenerate, as shown in (b). The energy cut $\Lambda$ near the Fermi level is $0.005$eV. To show the gap nodes on the FSs clearly, the range of the color-bar has been set to be the same.
\label{RPA}}
\end{figure}

First, we calculate  the bare  and RPA spin susceptibilities for BaCoSO at different doping levels as shown in Fig.\ref{susceptibility}.  The RPA spin susceptibility has a sharp peak near the wavevector $(\pi,\pi)$ in  both doping levels.
The peak mainly stems from the interaction between  the smaller hole pocket   near the $\Gamma$ point and the electron pocket near the $M$ point.  This situation is very similar to the case in iron-pnictides. The interaction between these two pockets are responsible for the superconducting pairing as well.  This is because the points on  the smaller hole pocket near the $\Gamma$ point and the electron pocket near the $M$ point contribute the largest density of states(DOS)  near the Fermi level. The pairing strength on the hole pocket attributed to the $d_{yz}$ orbital is always small because of its large band dispersion.

The  RPA results in the superconducting state are reported in Fig.\ref{RPA}. For the $0.6$ electron doping, similar to the mean field results, the leading superconducting instability turns out to have a d-wave like pairing symmetry, as shown in Fig.\ref{RPA}(a)(c).    In the $1.0$  heavy electron doping case, as shown in Fig.\ref{RPA}(b)(d) and (e), there are two leading superconducting instabilities which  are nearly degenerate. Both of them have many nodes on the FSs and the gap function is more complex compared to the $0.6$ electron doped case.  The superconducting pairing strength is also much weaker than those with the $0.6$ electron doping case.

There are significant differences between the RPA and mean field results. First, the superconducting gap on the two electron pockets near $Y$ point tends to have an uniform phase in the RPA results. Second,  the pairing strength on the FSs contributed by  the $d_{yz}$ orbital is much weaker in the RPA analysis than in the mean field approach.  Finally, competing superconducting pairing states is much easier to appear   in the RPA analysis as well.  These differences can be well understood in the build-in structure of the  RPA  analysis as the interaction between  the smaller hole pocket   near the $\Gamma$ point and the electron pocket near the $M$ point becomes dominant. Moreover, according to Eq.\ref{RPA_pair}, the large DOS leads to strong pairing strength, and to avoid repulsive interaction to save energy,  the pairings at these two areas tend to have a $\pi$ phase difference.

\begin{figure}
\centerline{\includegraphics[width=0.45\textwidth]{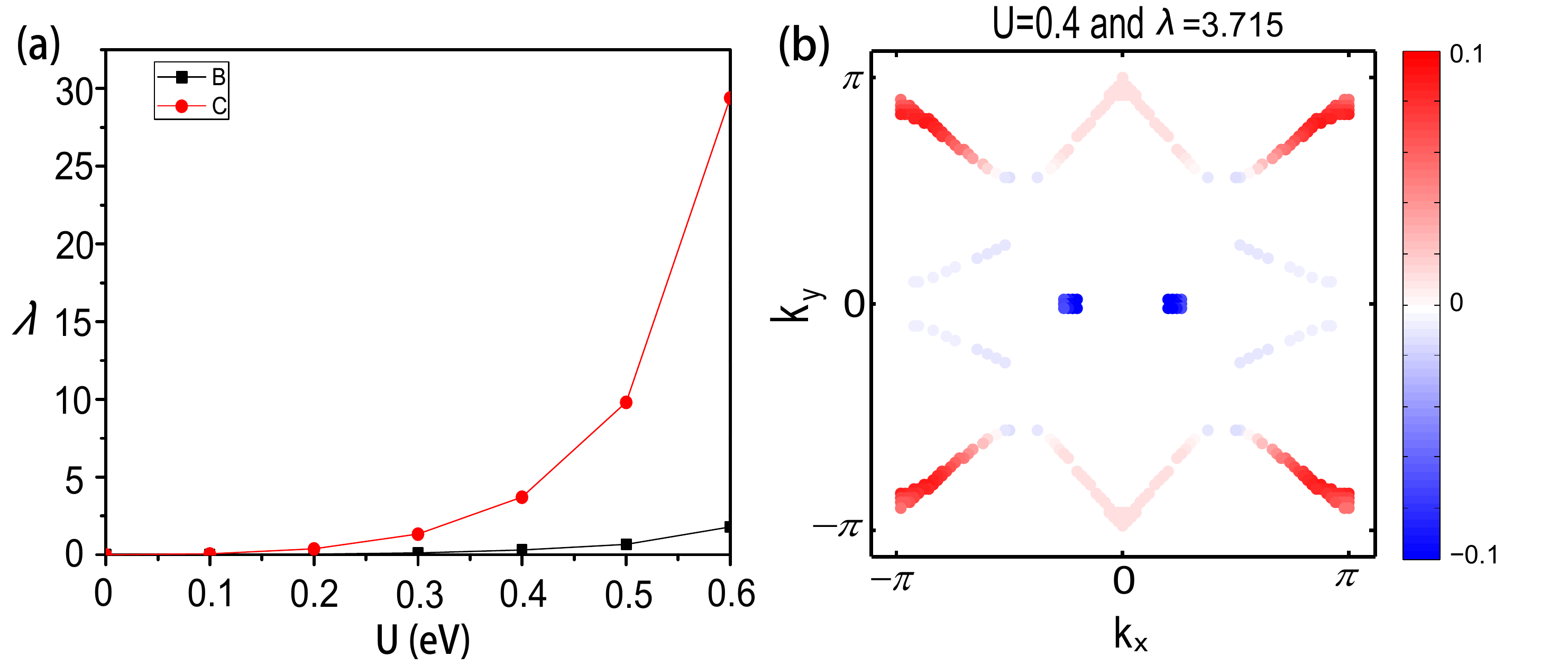}}
\caption{(color online) The leading pairing instability and superconducting gap on the FSs of BaCoSO for $J/U=0.2$ are shown in (a) and (b), respectively. Here, the electron doping level is about $1.0$ and the energy cut $\Lambda$ is $0.01$eV near the Fermi level.
\label{RPA_Ecut}}
\end{figure}

The importance of the interaction between the smaller hole pocket   near the $\Gamma$ point and the electron pocket near the $M$ point  in the RPA analysis can be demonstrated further.  By increasing the electron doping,  the smaller hole  pocket at $\gamma$ goes  a Lifshitz transition. At the $1.0$ doping, it  sinks just below Fermi level.  If the contribution from  the hole pocket is important,  the RPA calculation near the Lifshitz transition becomes very sensitive to the cutoff energy   $\Lambda$ from the Fermi level. For the results in Fig.\ref{RPA}(b)(d)(e), $\Lambda$ is taken to be $0.005$eV. If we increase this cutoff to involve the hole pocket contribution, the leading superconducting instability at the $1.0$ doping is expected to vary quickly. This expectation is demonstrated in  Fig.\ref{RPA_Ecut} in which  the cutoff energy is increased to  $0.01$eV.  It is clear that the leading superconducting instability  becomes similar to the $0.6$ electron doping case.

\section{summary and discussion}

In summary, we have carried out mean field and RPA calculation to analyze the possible superconducting ground state in the family of  materials similar to BaCoSO whose electronic structures are described by the three $t_{2g}$  orbitals.  It is found that a d-wave like superconducting state with gapless nodes is generally favored.

 The superconducting properties in this family of materials can help us to establish  fundamental principles regarding the emergence of superconductivity in unconventional high T$_c$ SCs.

 First, the energy scale of the parameters in this  family of materials are similar to those of iron-based superconductors. Therefore, the maximum T$_c$  that can be achieved here should be close to the maximum T$_c$  in the bulk material of iron-based superconductors if they share the same superconducting mechanism, which has been assumed in this paper.

 Second,  in cuprates and iron-based SCs,  the pairing symmetries are classified by the $D_{4h}$ group.  The superconducting states fall  into  specific irreducible representations of this high symmetry group.  It is difficult to mix different representations.  As a result,  a pure or close to a pure  d-wave  and  s-wave state have been realized in cuprates and iron-based supercnoductors respectively.   Here due to the absence of $C_4$ rotation system,  the superconducting state is classified by much lower symmetry group.  Thus, the superconducting state, in the term of the $D_{4h}$ group, is a mixture of s-wave and d-wave state.  Our results on the role of different orbitals, the location of gapless nodes and the pairing strength on different parts of Fermi surfaces  thus can provide critical information about the validity of theoretical methods and test different pairing mechanisms.

  Finally, the interaction between the smaller hole pocket   near the $\Gamma$ point and the electron pocket  near the $M$ point   is very similar to the case in iron-pnictides which are also characterized with the interaction between the hole pockets at the  $\Gamma$ point and the electron pockets at $M$ point\cite{iron_pnictide1,iron_pnictide2}.  Our results from the RPA analysis are consistent with those in iron-pnictides \cite{Kemper} . Both calculations suggest  that the interactions are responsible for superconductivity. However,  in iron-chalcogenides\cite{ding,feng,zhou},   the simple RPA result has been seriously challenged because the high T$_c$ superconductivity can still be achieved in the absence of hole pockets.  As the hole pockets can also sink below Fermi level by doping in this family of materials,  the validity of the RPA analysis can be further tested.  \textbf{For example,  the heavy electron doping may be achieved by substituting Co with Ni atoms.} We want to mention that strong superconducting instability cannot  be obtained  by  the standard functional renormalization group (FRG) method \cite{FRG1,FRG2,FRG3} in the above model.  Combining all these results and  the fact that the FRG  is also only valid in the weak interaction region,  observing high $T_c$ superconductivity in the family of materials may finally explore the limitation of these standard approaches.

\section{acknowledgement}
This work is supported by the Ministry of Science and Technology of China 973 program(Grant No. 2015CB921300), National Science Foundation of China (Grant No. NSFC-1190020, 11534014, 11334012), and   the Strategic Priority Research Program of  CAS (Grant No. XDB07000000).

\end{document}